\documentclass[aps,prb,reprint,groupedaddress]{revtex4-2}

\usepackage{amsmath}
\usepackage{amssymb}
\usepackage{mathrsfs}

\usepackage{graphicx}
\usepackage{txfonts}

\begin{document}

\title{Diffuse scattering on Ising chain with competing interactions}

\author{F.A. Kassan-Ogly}
\email{Felix.Kassan-Ogly@imp.uran.ru}
\affiliation{M.N. Mikheev Institute of Metal Physics of Ural Branch of Russian Academy of Sciences, 620108 Ekaterinburg, Russia}

\author{A.V. Zarubin}
\email{Alexander.Zarubin@imp.uran.ru}
\affiliation{M.N. Mikheev Institute of Metal Physics of Ural Branch of Russian Academy of Sciences, 620108 Ekaterinburg, Russia}

\author{A.I. Proshkin}
\affiliation{M.N. Mikheev Institute of Metal Physics of Ural Branch of Russian Academy of Sciences, 620108 Ekaterinburg, Russia}

%\date{\today}

\begin{abstract}
We considered the Ising 1D chain in an external magnetic field taking into account the nearest and next-nearest neighbor interactions. By the method of Kramers--Wannier transfer-matrix, the rigorous analytical expression for Fourier-transform of pair spin-spin correlation function was obtained, and the temperature evolution of the scattering was analyzed for various relations of exchange parameters.
\end{abstract}


\maketitle

\section*{Introduction}

In the last decades~\cite{Giamarchi:2003}, the low-dimensional magnets with special properties, promissory for high-tech industry~\cite{Roth:2004}, have obtained advanced development. The elastic magnetic neutron scattering is the sole source of knowledge on the magnetic structure arrangement inside a material. For the interpretation of diffraction patterns, it is very important to calculate the pair spin-spin correlation function. Unfortunately, up to now, there is lack of exact solutions for correlation function in 3D objects. Nevertheless, there exist 1D and 2D exactly solvable models that help to understand the nature and peculiar properties of magnetic materials. 

The most convenient and widely used for such purposes is the Ising model that gives qualitative description and allows detecting special features of magnetic materials, inaccessible within the framework of various approximate approaches~\cite{Takahashi:1999}. Of course, the Ising model has several well-known solutions~\cite{Baxter:1982}, but systematic research of scattering problems was not carried out~\cite{Kassan-Ogly:2001}.

\section{Fourier transform of pair spin-spin correlation}

The derivation of elastic magnetic scattering of unpolarized neutrons on 1D monatomic equidistant chain of spins in the Ising model with allowance for the nearest and next-nearest neighbor interactions with the Hamiltonian
\begin{equation}
\mathscr{H}=-J\sum_{r}(\boldsymbol{\sigma}_{r},\boldsymbol{\sigma}_{r+1})-J'\sum_{r}(\boldsymbol{\sigma}_{r},\boldsymbol{\sigma}_{r+2})-\sum_{r}(\mathbf{h},\boldsymbol{\sigma}_{r})
\label{eq:H}
\end{equation}
reduces to the computation of Fourier transform of pair spin-spin correlation function
\begin{equation}
K(q)=\frac{1}{N}\sum_{r,r'}\sum_{\{\sigma\}}\frac{e^{-\beta\mathscr{H}}(\boldsymbol{\sigma}_{r},\boldsymbol{\sigma}_{r'})}{Z}e^{-iqa(r-r')},
\label{eq:KF:0}
\end{equation}
where $q$ is the scattering wave-vector, $J$ is the exchange interaction parameter, $\boldsymbol{\sigma}_{r}$ are spin operators, that take the values $+1$ or $-1$, subscripts of operators $\boldsymbol{\sigma}_{r}$ run over the lattice sites, $\mathbf{h}$ is an external magnetic field, $T$ is the temperature, $1/\beta=k_{\text{B}}T$, $k_{\text{B}}$ is Boltzmann constant, (hereinafter we omit it), $a$ is the lattice spacing, the sum over $\{\sigma\}$ stands for all possible spin configurations, $N$ is the number of lattice sites, $Z$ is the partition function of the system.

In the method of Kramers–Wannier transfer-matrix the Fourier transform of pair spin-spin correlation function takes on a form
\begin{equation}
K(q_{l})=\frac{1}{N}\sum_{r=0}^{N}\sum_{r'=0}^{N}\frac{\operatorname{Tr}(\mathbf{V}_{1}\mathbf{V}_{2}\ldots\boldsymbol{\sigma}_{r}\mathbf{V}_{r}\ldots\boldsymbol{\sigma}_{r'}\mathbf{V}_{r'}\ldots\mathbf{V}_{N})}{\operatorname{Tr}(\mathbf{V}_{1}\mathbf{V}_{2}\ldots\mathbf{V}_{N})}e^{-iqa(r-r')},
\label{eq:KF:1}
\end{equation}
where $\mathbf{V}_{r}$ is the transfer-matrix~\cite{Baxter:1982}.

With allowance for the nearest and next-nearest neighbor interactions in an external magnetic field the transfer-matrix has the secular equation~\cite{Kassan-Ogly:2001}
\begin{multline}
\lambda^{4}-2e^{\beta(J+J')}\cosh(\beta h)\lambda^{3}+2e^{\beta J'}\sinh(2\beta J)\lambda^{2} \\
-4e^{\beta(J+J')}\sinh(2\beta J')\cosh(\beta h)\lambda-4\sinh^{2}(2\beta J')=0,
\label{eq:SEQ}
\end{multline}
and the Fourier transform of correlation function (Eq.~\ref{eq:KF:1}) is composed of two terms:
\begin{equation}
K(q)=M^{2}L(q)+D(q),
\label{eq:KF:2}
\end{equation}
where $L(q)$ is the Laue function that positions the Bragg reflections, and $D(q)$ is the function that determines the diffuse scattering:
\begin{equation}
L(q)=\frac{1}{N}\frac{\sin^{2}(qaN/2)}{\sin^{2}(qa/2)},
\label{eq:LF}
\end{equation}
\begin{equation}
K(q)=\sum_{i=2}^{4}\nu_{i}\frac{1-\Lambda_{i}^{2}}{1-2\Lambda_{i}\cos(qa)+\Lambda_{i}^{2}},
\label{eq:DF}
\end{equation}
\[
M=\frac{T}{\lambda_{1}}\frac{\partial\lambda_{1}}{\partial h},\quad\Lambda_{i}=\frac{\lambda_{i}}{\lambda_{1}},\quad\sum_{i=2}^{4}\nu_{i}=1-M^{2}.
 \]
Here $M$ is the magnetization, $\lambda_{i}$ are the roots of the transfer-matrix secular equation, $\lambda_{1}$ is the maximum eigenvalue of the transfer-matrix, and $\nu_{i}$ are cumbersome expressions that do not depend on the scattering wave-vector~$q$.

The function $L(q)$ determines the Bragg scattering that is specified by the lattice parameter and does not depend on the sign of exchange interaction. The presence of magnetic field complicates the calculation of scattering intensity by inducing the magnetization. The whole temperature dependence of the Fourier transform of correlation function is contained in the parameters $\Lambda_{i}$. Thus for studying the temperature behavior of scattering it is sufficient to only treat the function of diffuse scattering~$D(q)$.

\begin{figure}[htb]
\begin{center}
\includegraphics[width=0.8\columnwidth]{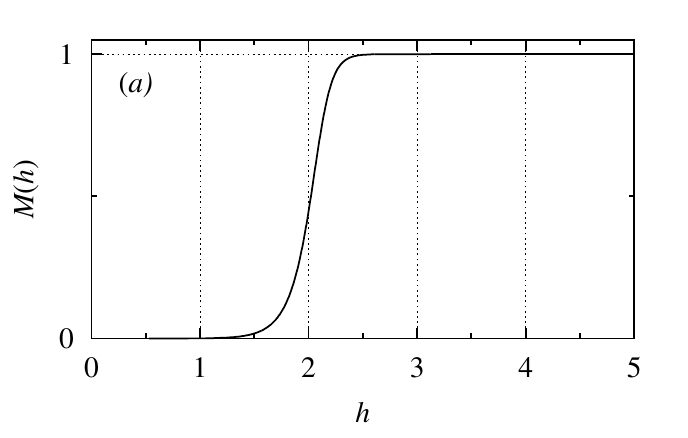}\par
\includegraphics[width=0.8\columnwidth]{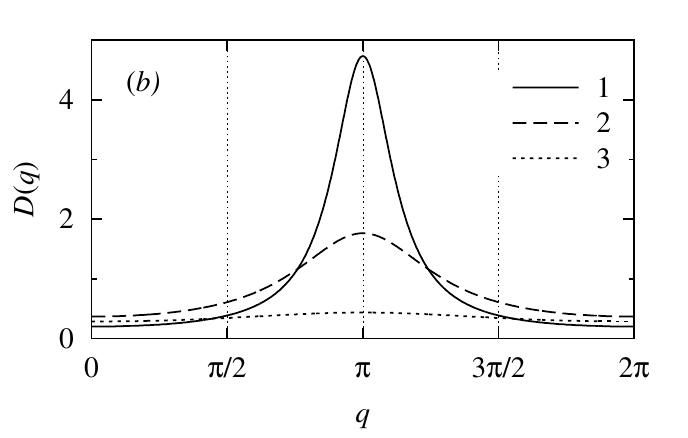}
\end{center}
\caption{(a) Magnetization of 1D chain at $J=-1$ and $T=0.15$; (b) The $D(q)$ dependence at $J=-1$, $h=1<h_{\text{c}}$ (curve~1), $h=2=h_{\text{c}}$ (curve~2) and $h=3>h_{\text{c}}$ (curve~3), $T=1$}
\label{fig:1}
\end{figure}

Let us consider the scattering on the Ising chain with both negative exchange interactions in an external field. In the case of antiferromagnetic nearest neighbor interaction ($J<0$, $J'=0$) (Fig.~\ref{fig:1}) there appears the frustration field ($h_{\text{c}}=-2J$), at which the system behavior cardinally alters. Above this field, the configurations of $\{+-+-+-\}$ type become energetically disadvantageous, and the whole system of spins aligns along the direction of an external field with a sole energetically advantageous configuration $\{++++++\}$ (Fig.~\ref{fig:1}a).

In the fields below the frustration one, the diffuse scattering concentrates at low temperatures at the points precisely between the Bragg positions $q=2\pi(2n+1)/2a$ ($n=0, \pm 1, \pm 2, \ldots $). Above the frustration field, (even at $T\to 0$) the diffuse scattering takes the shape of broad peaks of low intensity that do not transform into delta-functions (Fig.~\ref{fig:1}b).

In the case of competing interactions between nearest and next-nearest neighbors ($J<0$, $J'<0$) there appear two frustration fields: $h_{\text{c}1}=-2(J-2J')$ and $h_{\text{c}2}=-2(J+J')$.

\begin{figure}[htb]
\begin{center}
\includegraphics[width=0.8\columnwidth]{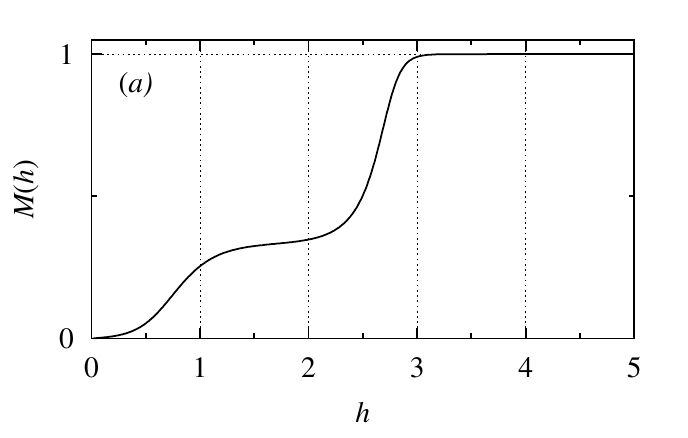}\par
\includegraphics[width=0.8\columnwidth]{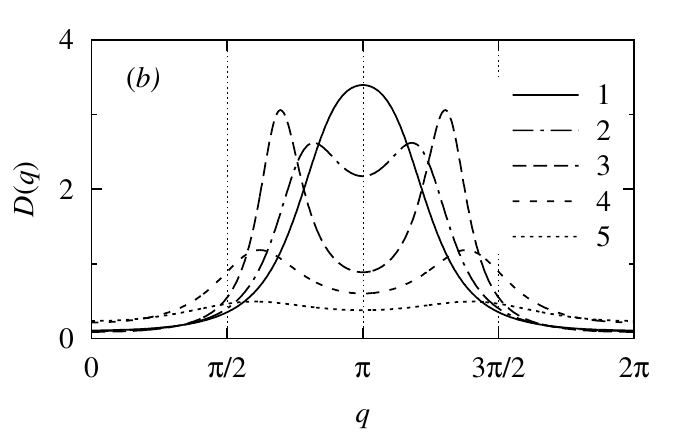}
\end{center}
\caption{(a) Magnetization of 1D chain at $J=-1$, $J'=-0.3$ and $T=0.15$; (b) The $D(q)$ dependence at $J=-1$ and $J'=-0.3$, at $h=0.5<h_{\text{c}1}$ (curve~1), $h=0.8=h_{\text{c}1}$ (curve~2), $h=1.5<h_{\text{c}2}$ (curve~3), $h=2.6=h_{\text{c}2}$ (curve~4) and $h=3<h_{\text{c}2}$ (curve~5), $T=0.5$}
\label{fig:2}
\end{figure}

Below the lower frustration field, the energetically advantageous configurations of $\{+-+-+-\}$ type are replaced by configurations of $\{++-++-\}$ type in the interval between the frustration fields, and above the upper frustration field, $h>h_{\text{c}2}$ the spin system aligns along an external field $\{++++++\}$. Figure~\ref{fig:2} exemplifies the magnetization and diffuse scattering in such cases.

\begin{figure}[htb]
\begin{center}
\includegraphics[width=0.8\columnwidth]{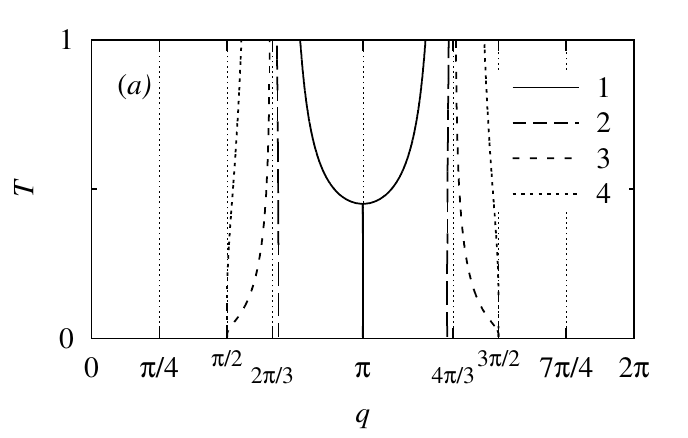}\par
\includegraphics[width=0.8\columnwidth]{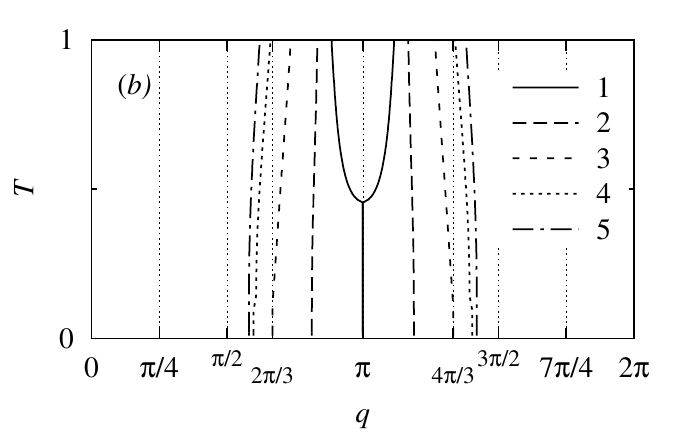}
\end{center}
\caption{(a) Temperature dependence of the wave-vector peak position at $h=0$ and: $J=-1$, $J'=-0.4$ (curve~1), $J=-1$, $J'=-0.5$ (curve~2), $J=-1$, $J'=-0.55$ (curve~3), $J=-1$, $J'=-1$ (curve~4); (b) Temperature dependence of the wave-vector peak position at $J=-1$, $J'=-0.3$ and $h=0.6$ (curve~1), $h=0.8$ (curve~2), $h=1.5$ (curve~3), $h=2.6$ (curve~4) and $h=4$ (curve~5)}
\label{fig:3}
\end{figure}

The behavior of diffuse scattering has peculiar features at different relations of the model parameters. In a pure antiferromagnetic case ($J<0$, $J'=0$), the Fourier transform of correlation function peaks adopt constant positions $q_0=(2n+1)\pi $ ($n=0, \pm 1, \pm 2, \ldots $). In the absence of an external magnetic field, but with exchange interactions of arbitrary values and signs the peak positions of the scattering wave-vector (Fig.~\ref{fig:3}a) are determined by the general formula
\begin{equation}
q_{0}=\arccos\left(-\frac{1}{2}\frac{\sinh(\beta J)}{\sinh(2\beta J')}e^{2\beta J'}\sqrt{\sinh^{2}(\beta J)+e^{-4\beta J'}}\right)\pm2\pi n.
\label{eq:Q0:2}
\end{equation}

The formula is obtained from the expression for Fourier transform of pair spin-spin correlation function (equation~(24) in Ref.~\onlinecite{Kassan-Ogly:1989:}). It should be noted that Stephenson~\cite{Stephenson:1970} had been the first who obtained the expression for correlations in direct space, but without derivation of Fourier transform of correlation function.

The behavior of the diffuse scattering peaks with allowance of an external magnetic field is shown in Fig.~\ref{fig:3}b.

\section*{Summary}

An external magnetic field generates two effects on the scattering. First, it creates an additional term, namely, Bragg reflections, modulated by the magnetization squared. Second, magnetic field changes the intensity of diffuse scattering and the positions of scattering wave-vector.

Depending on the values and signs of exchange interaction, and the value of magnetic field, there appear the following scenarios of the diffuse scattering behavior.

1. At negative nearest-neighbor interaction and weak next-nearest neighbor interaction at a field below the lower frustration one, the diffuse scattering at $T\to 0$ accumulates in delta-functions in between Bragg reflections, i.e. the phase transition appears as in pure antiferromagnetic case. 

2. In the case of moderate next-nearest neighbor interaction, at high temperatures the broad diffuse peaks arise in incommensurate positions. Then these peaks pairwise approach and merge at some temperature (the lock-in transition that depends on the value of magnetic field) (Fig.~\ref{fig:3}, curve~1). 

3. At strong next-nearest neighbor interaction and magnetic field between the lower and upper frustration fields, the broad diffuse peaks also arise in incommensurate positions. Then these peaks grow, changing simultaneously their positions tending at $T\to 0$ to $q_0=2\pi/3+2n\pi$ ($n=0, \pm 1, \pm 2, \ldots $) (Fig.~\ref{fig:3}) and acquire the delta-function shape.

4. Above the upper frustration field, the diffuse peaks grow and change their positions acquiring the delta-function shape at $T\to 0$ and the Bragg positions.

5. At all frustration fields, the diffuse scattering remains smooth up to the zero temperature without phase transition (frustration regimes) (Fig.~\ref{fig:3}, curve~2).

\bigskip

The research was carried out within the state assignment of FASO of Russia (theme “Quantum” No. 01201463332).

\bibliographystyle{apsrev4-2}
\bibliography{corrfunc2015_arxiv}

\end{document}